# Mechanical Characterization of Amyloid Fibrils Using Coarse-Grained Normal Mode Analysis

Gwonchan Yoon[1], Jinhak Kwak[1], Jae In Kim[1], Sungsoo Na[2], and Kilho Eom[2]

*Department of Mechanical Engineering, Korea University, Seoul 136-701, Republic of Korea*

[1]These authors (G.Y., J.K., and J.I.K.) made equal contribution to this work.

[2]Correspondence should be addressed to S.N. (E-mail: nass@korea.ac.kr) and K.E. (E-mail: kilhoeom@korea.ac.kr or kilhoeom@gmail.com).

**Abstract**

Recent experimental studies have shown that amyloid fibril formed by aggregation of β peptide exhibits excellent mechanical properties comparable to other protein materials such as actin filaments and microtubules. These excellent mechanical properties of amyloid fibrils are related to their functional role in disease expression. This indicates the necessity to understand how an amyloid fibril achieves the remarkable mechanical properties through self-aggregation with structural hierarchy, highlighting the structure-property-function relationship for amyloids, whereas such relationship still remains elusive. In this work, we have studied the mechanical properties of human islet amyloid polypeptide (hIAPP) with respect to its structural hierarchies and structural shapes by coarse-grained normal mode analysis. Our simulation shows that hIAPP fibril can achieve the excellent bending rigidity via specific aggregation pattern such as antiparallel stacking of β peptides. Moreover, we have found the length-dependent mechanical properties of amyloids. This length-dependent property has been elucidated from Timoshenko beam model that takes into account the shear effect on the bending of amyloids. In summary, our study sheds light on the importance of not only the molecular architecture, which encodes the mechanical properties of the fibril, but also the shear effect on the mechanical (bending) behavior of the fibril.

## I. Introduction

In the past decade, it was found that some denatured proteins play a significant role in disease expressions in such a way that such proteins act as a catalyst that assists the forming of long protein fibrils [1-3]. For instance, Alzheimer's disease is expressed due to presence of undesired formation of a long amyloid fibril that originates from the aggregation of amyloid β peptide, that is, self-assembly and self-propagation of degenerated proteins in the human brain [4, 5]. A long amyloid fibril was also found for the pancreas of the type II diabetes, and this amyloid fibril is named as human islet amyloid polypeptide (hIAPP), which is a key factor that involves in the insulin-secreting inhibition [6]. In particular, hIAPP$_{20-29}$(SNNFGAILS) is a molecular chain that is responsible for the formation of amyloidogenic core that serves as a catalyst in the assembly of fibrils that can replace insulin-secreting β-cells at the islet of pancreas. Recently, there have been current attempts [7, 8] to unveil the supramolecular structures of amyloid fibril such as hIAPP$_{20-29}$ fibrils, in order to gain fundamental insights into the formation of amyloidogenic core, by using solid-state NMR [9]. Moreover, there have been recent efforts [10-14] to theoretically provide the possible molecular architecture of amyloid fibrils. Specifically, a recent study [15] suggests eight possible molecular structures for amyloid fibrils based on the hierarchy of multiple ladder-shaped β sheets (i.e. cross-β amyloid).

Amyloid fibrils that are formed due to aggregation of β sheet are mechanically stable in the physiological conditions, so that they can stably replace the specific cells (e.g. insulin-secreting β-cell) at the specific site (e.g. islet of pancreas). This hypothesis has led researchers to study not only the mechanisms of protein aggregation but also the mechanical stability of amyloid fibrils. A recent study by Knowles et al. [16] has remarkably elucidated that a variety of amyloid fibrils, which are composed of mechanically weak proteins, exhibits the excellent mechanical properties such that the elastic stiffness (i.e. bending modulus) of amyloid fibrils is comparable to other protein materials such as actin filament and microtubule. This sheds light on the mechanical stability of amyloid fibrils that consists of even weak proteins. In a recent study [17], the bending modulus of β-lactoglobulin fibrils was measured based on polymer theory coupled with atomic force microscope (AFM)-based imaging technique. It is shown that the mechanical properties of β-lactoglobulin fibrils are highly correlated with structural hierarchy and assembly, particularly helical periodicity, fibril length, and fibril thickness. The experimental observations, which support the hypothesis on the mechanical stability of amyloid fibril, are insufficient to elucidate the physical origin of the mechanical stability of fibrils. Recently, Buehler and coworkers [18] have interestingly found, by using atomistic simulations, that the geometrical confinement of β-sheet results in the enhancement of the mechanical properties of protein crystal. This indicates that the hydrogen bonds between β-sheet layers act as a chemical glue between the layers, which increases the mechanical stability of a protein crystal. This is consistent with the hypothesis by Knowles et al. [16], who suggested that intermolecular forces between β-sheet layers is a key parameter that determines the mechanical properties of amyloid fibril. This sheds light on the structure-property-function relation for protein materials, which implies that the mechanical properties of protein materials are encoded in the structural hierarchy and assembly.

Mechanical characterization of protein materials has been made possible due to recent experimental and computational techniques. For instance, AFM-based nanoindentation and force spectroscopy has recently been highlighted for measurement of the mechanical properties of soft protein materials such as viral capsid [19, 20], microtubule [21], spider silk protein [22], amyloid fibril [16, 23], and so forth. Despite its ability to straightforwardly provide the mechanical properties of protein materials based on continuum mechanics theory and/or polymer theory, AFM-based experiments require the cautious sample preparation that may significantly affect the measured properties. Specifically, as described in Ref. [24], the molecular structures of amyloid fibril may critically depend on the sample preparation, indicating that the mechanical properties of amyloid fibril may be also governed by the sample preparation. In addition, Kis et al. [21] reported that the mechanical properties (e.g. bending rigidity) of microtubule is strongly dependent on the experimental conditions such as temperature. Moreover, AFM nanoindentation is insufficient to unveil the details of the molecular structure of amyloid fibrils, which implies the limitations in understanding the structure-property relationship.

One of possible routes to understand the structure-property relationship is to employ the computational simulations based on atomistic models. As described above, molecular dynamics (MD) simulation has enabled the characterization of mechanical properties of protein crystal with respect to its molecular architecture [16, 18]. However, MD simulations are still computationally restrictive for studying the dynamics of the macromolecular structure that requires a simulation at the large spatial and temporal scales [25, 26]. In a recent decade, in order to resolve the computational restrictions encountered in MD simulation, there has been a recent attempt to develop the coarse-grained models that are developed in such a way that degrees of freedom for molecular system is reduced as well as the potential field ascribed to the system is also simplified [27-31]. One of intriguing coarse-grained models to describe the protein structure is the elastic network model (ENM), which was first suggested by Tirion [32] and later by several researchers [33-37]. The key feature of ENM is to describe the protein structure based on Cα atoms in such a way that Cα atoms within the neighborhood are connected by an elastic spring with an identical force constant. Despite its simplicity, ENM is very robust in predicting the conformational dynamics of protein structures. This robustness is attributed to the principle that protein dynamics and/or mechanics are encoded in the protein structure, namely native topology [38-40]. This implies that ENM is suitable to depict the protein structure that encodes its mechanical properties. This has several researchers to employ ENM for studying the mechanical properties of protein fibril structures. For instance, Yoon et al. [41] have developed the mesoscopic model of protein crystals by using Gō-like model and/or ENM in order to characterize the mechanical properties of protein crystals. Normal Mode Analysis (NMA) with ENM has recently been employed to study the molecular mechanics of the cytoskeletal crosslinker, i.e. α-actinin rod domain, and the simulation results of NMA are compared with those computed from force-induced MD for bending, extension and twisting [42]. Recently, Buehler and coworkers [43] have studied the mechanical properties of amyloid fibrils composed of A$\beta_{1-40}$ by using NMA with ENM. Their study [43] shows that molecular architecture does play a key role in the mechanical properties of amyloid fibril. A recent study by Deriu et al. [44] have provided the theoretical model of microtubule based on ENM in order to gain the fundamental insights into its mechanical properties. These studies [41-44] shed light on the ENM for studying the mechanical properties of biological macromolecules and/or assemblies.

In this work, we have studied the mechanical properties of hIAPP fibril that plays a functional role in the expression of type II diabetes. Our mechanical characterization is performed by the computational modeling techniques based on NMA with ENM. Here, we have considered the possible molecular architectures of hIAPP as suggested in Ref. [15]. In particular, recent experimental studies [9, 24] suggest that hIAPP$_{20-29}$ can exist in two major different forms: antiparallel hetero (AHT) configuration and/or parallel homo (PHM) configurations (for more details, see Fig. 1). We have characterized the mechanical properties of hIAPP with respect to its molecular architectures (i.e. configurations) as shown in Fig. 1. Moreover, we have found that the mechanical properties of hIAPP fibrils depend on their fibril length such that when the fibril length reaches ~70 nm, the fibril becomes to be mechanically stable. In addition, we have also showed that twisted configuration for the interface between β-sheet ladders allows the fibril to exhibit the isotropic bending rigidity. Remarkably, our finding of the length-dependent elastic properties (i.e. bending rigidity) for amyloid fibril is consistent with the theoretical continuum mechanics model, which shows that bending properties of fibril are also attributed to the shear deformation. To our best knowledge, since there have been few attempts to characterize the mechanical properties of hIAPP fibrils, our study may provide the milestone for understanding the mechanical stability of hIAPP fibrils with respect to their property-encoded molecular architectures.

## II. Theory and Model

### 1. Continuum theory

We have employed the continuum mechanics theory in order to extract the mechanical properties of hIAPP fibril from its vibrational behavior that is predicted from coarse-grained model (i.e. NMA with ENM). Since a fibril exhibits the characteristic dimension such that the longitudinal dimension (i.e. length) of a fibril is much larger than other transverse dimensions (i.e. cross-sectional dimension), which implies that a fibril can be modeled as a one-dimensional elastic beam. The deformation of an elastic beam can be described by three major modes: bending, stretching, and twist. The mechanical properties (i.e. bending rigidity, axial stiffness, and torsional rigidity) corresponding to these three dominant deformation modes can be estimated from the vibrational characteristics that can be measured from atomistic (or coarse-grained) model. In order to relate the measured resonance behavior of a fibril to its mechanical properties, we utilize the Euler-Bernoulli beam model, which is widely used for analysis on the mechanical behavior of beam-like nanostructures such as biological filament [45] and/or nanotubes [46, 47]. The details of Euler-Bernoulli beam model are presented in the Supporting Information.

## 2. Coarse-Grained Model: Measurement of Natural Frequencies for Fibrils Based on Molecular Structures

### A. Construction of the hIAPP fibril

The human islet amyloid polypeptide (hIAPP) fibril consists of a building block composed of four identical β sheets whose sequence is given as "NFGAILS". Specifically, the hIAPP fibril can be constructed in such a way that a building block is periodically repeated in the longitudinal direction (i.e. fibril axis) of a fibril. The molecular structure of a building block for hIAPP fibril is deposited in protein data bank (PDB) with PDB code of 2KIB, which was identified by standard 2D NMR spectra [9]. Since a recent studies [24] report that there could be other molecular structure for the building block, we consider the four possible molecular architectures for the building block of hIAPP fibril. In particular, as shown in Fig. 1, we denote four β sheets as $\beta_1$, $\beta_2$, $\beta_3$, and $\beta_4$, respectively, which composes the hIAPP fibril. The structure deposited as 2KIB in PDB described the antiparallel hetero (AHP) configuration such that $\beta_1$ and $\beta_3$ forms the registered antiparallel strand, and that the position vector of other registered sheets (i.e. $\beta_2$ and $\beta_4$) can be obtained from rotating the registered antiparallel strand (i.e. formed by $\beta_1$ and $\beta_3$) followed by the translation along the direction perpendicular to the surface formed by the registered antiparallel strand by the amount of the distance between $\beta_1$ and $\beta_2$ (or $\beta_3$ and $\beta_4$) [see Fig. 1d]. In the similar manner, we have considered the four possible configurations for the building block of hIAPP fibril: (i) parallel homo (PHM), (ii) parallel hetero (PHT), (iii) antiparallel homo (AHM), and (iv) antiparallel hetero (AHT) (for details, see Fig. 1). Based on the experimental observation [9] of the molecular structure for hIAPP fibril, it is presumed that hIAPP$_{20-29}$ fibril exhibits the twisted structure along the fibril axis. Specifically, it was shown that the helical pitch for hIAPP fibril is 25.811 nm, and that 72 β sheets can form one helical pitch of the fibril [9]. Based on this, we have constructed the fibril in such a way that the building block (in a specific configuration) is repeated but rotated 20° about the fibril axis. The constructed model of hIAPP fibril is shown in Fig. 2.

B. Elastic Network Model (ENM)

For long amyloid fibril, the atomistic model is computationally restrictive because of large degrees of freedom and complex potential field. In order to compute the natural frequencies for the deformation modes (e.g. bending mode, torsional mode, and axial stretching mode), we have employed the coarse-grained structural model that enables the computationally efficient analysis on the vibrational properties of a fibril. Specifically, we have utilized the elastic network model (ENM) [32-37] that has been widely taken into account for studying the low-frequency motions of proteins. Here, the molecular structure of an amyloid fibril is described by ENM that regards the molecular structure as the harmonic spring network for Cα atoms. In particular, ENM is constructed in such a way that Cα atoms within the neighborhood are connected by an elastic (harmonic) spring with identical force constant. The robustness of ENM in predicting the dynamic behavior of protein structures is attributed to the principle that the dynamic characteristic is encoded in the native topology of a protein [28, 48]. The potential energy, $V$, for ENM is given by,

$$V = \frac{\gamma}{2} \sum_{i=1}^{N} \sum_{j \neq i}^{N} \left[ \left| \mathbf{r}_i - \mathbf{r}_j \right| - \left| \mathbf{r}_i^0 - \mathbf{r}_j^0 \right| \right]^2 \cdot H\left(r_c - \left| \mathbf{r}_i^0 - \mathbf{r}_j^0 \right|\right) \qquad (1)$$

where $\gamma$ is a force constant of a harmonic spring, $N$ is the total number of residues, $r_i$ is the coordinates of $i$-th residue (alpha carbon atom), $r_c$ is a cut-off distance, $H(x)$ is the Heaviside unit step function defined as $H(x) = 0$ if $x < 0$; otherwise $H(x) = 1$, and superscript 0 indicates the equilibrium conformational state. Here, we set the parameters such as $\gamma = 0.5$ kcal·mol$^{-1}$Å$^{-2}$ for non-bonded interaction (e.g. see Ref. [36]) and $\gamma = 100$ kcal·mol$^{-1}$Å$^{-2}$ for the covalent bond stretch, and $r_c = 12$ Å.

For implementation of NMA, we need to compute the stiffness matrix based on a potential energy given by Eq. (1). The stiffness matrix $\mathbf{K}$ is composed of $3 \times 3$ block matrices $\mathbf{K}_{ij}$, which is defined as $\mathbf{K}_{ij} = -\partial^2 V / \partial \mathbf{r}_i \partial \mathbf{r}_j$ [49].

$$\mathbf{K}_{ij} = -\left[ \gamma H\left(r_c - \left| \mathbf{r}_i^0 - \mathbf{r}_j^0 \right|\right) \frac{\left(\mathbf{r}_i^0 - \mathbf{r}_j^0\right)^T \left(\mathbf{r}_i^0 - \mathbf{r}_j^0\right)}{\left| \mathbf{r}_i^0 - \mathbf{r}_j^0 \right|^2} \right] \times \left(1 - \delta_{ij}\right) - \delta_{ij} \sum_{r \neq i}^{N} \mathbf{K}_{ir} \qquad (2)$$

Here, $\delta_{ij}$ is the Kronecker delta defined as $\delta_{ij} = 1$ if $i = j$; otherwise $\delta_{ij} = 0$, and $v^T$ represents the transpose of vector $v$. Based on the stiffness matrix **K**, the vibration of an amyloid fibril can be analyzed by NMA such as $\mathbf{Kq} = m\omega^2\mathbf{q}$, where $m$ is the atomic mass of a Cα atom, and $\omega$ and **q** represent the natural frequency and its corresponding eigenmode, respectively. It should be noted that we exclude the six 0 modes that correspond to the rigid body motions, i.e. translational and rotational rigid body motions. In order to visualize the vibrational deformation of a fibril, we displace the fibril in such a way that each Cα atom is moved along the eigenmode: The position vector after vibrational deformation, $\mathbf{R}^*$, is given as $\mathbf{R}_i^* = \mathbf{r}_i^0 + \alpha_i \mathbf{u}_i^k$, where $\mathbf{r}_i^0$ is the position vector for $i$-th Cα atom for the fibril structure in undeformed configuration (i.e. native conformation), $\mathbf{u}_i^k$ represents the directional vector for $i$-th Cα atom at $k$-th normal mode, and $\alpha_i$ is a constant that is determined in such a way that the root mean-square distance (RMSD) between deformed and undeformed configurations becomes 1 nm.

## III. Results

*1. Vibrational Modes of a hIAPP Fibril*

First, in order to ensure whether the vibrational motion of a hIAPP fibril is well dictated by a continuum mechanics model, we consider the deformation modes of hIAPP fibril, which were computationally obtained from a coarse-grained model (i.e. NMA with ENM). Fig. 2 depicts the major vibrational mode shapes for a hIAPP fibril (in the configuration of AHT) that exhibits the length of 17.045 nm (equivalent to a single pitch). As shown in Fig. 2, these vibrational modes correspond to the bending, torsional, and axial stretching deformations, respectively, which suggests that the vibrational deformation can be analyzed by a continuum mechanics model described in Section II.1. Here, the vibrational modes for rigid body motions (six 0 modes) are excluded. The lowest (i.e. $7^{th}$ modes) frequencies are ~0.1 THz and its corresponding vibrational mode is bending mode, which is comparable to the vibration behavior of Aβ amyloid fibril reported in Ref. [43], We have two fundamental bending modes due to anisotropic property of cross-sectional area. Specifically, there are two major principal cross-sectional moments of inertia such as $I_{max}$ and $I_{min}$, which represents the cross-sectional moment of inertia with respect to two principal neutral axes, respectively. As shown in Fig. 2a, the two lowest frequency normal modes corresponding to bending deformation modes are well dictated by a continuum mechanics model depicted in Eq. (S3.a) with amplitudes of $A_1^B = 0.03$ (for bending about soft axis) and $A_2^B = 0.01$ (for bending about stiff axis). Because the thermal fluctuation is mostly attributed to the low-frequency motions [32, 50], the thermal fluctuation behavior of a hIAPP fibril is described by the bending fluctuation. This is consistent with AFM image-based observation that shows the bending fluctuation motion of an amyloid fibril [17]. Moreover, imaging-based experiment [51] reports that the fluctuation of a biological fibril (e.g. microtubule) is well dictated by a continuum elastic beam model. In addition, the bending deformation is a key factor that determines the mechanical behavior of the fibril in the viewpoint of statistical mechanics, particularly the worm-like chain model [52, 53] presuming that the fluctuation of a biological fiber (e.g. DNA [54-56], microtubule [51], amyloid fibril [17], etc.) is ascribed to the fluctuation of bending angles. On the other hand, the high-frequency normal modes, which rarely contribute to the thermal fluctuation, are well depicted by axial stretching and torsional modes (see Fig. 2b). This clearly elucidates that the fundamental mechanics of a hIAPP fibril can be well represented by the bending deformation. Moreover, the high-frequency normal modes also include the coupled deformation modes such as coupling between axial stretching and torsional modes. In our study, we discard the high-frequency deformation modes corresponding to the coupled modes. In summary, the vibrational deformation modes of a hIAPP fibril, which are computationally obtained from molecular structure-based coarse-grained model, can be depicted by a continuum mechanics theory such as Euler-Bernoulli beam theory.

*2. Mechanical Properties of a hIAPP Fibril*

We have measured the mechanical properties (i.e. bending rigidity, axial stiffness, and torsional rigidity) of hIAPP fibrils based on the natural frequencies (for a specific deformation, e.g. bending deformation, etc.) that were evaluated from coarse-grained model. In particular, the mechanical properties are obtained from the relationship between the elastic stiffness and the natural frequency, depicted in Eq. (S4). Fig. 3 shows the elastic moduli (e.g. bending modulus, torsional modulus, and axial stiffness) of a hIAPP fibril in the AHT configuration as a function of the fibril length. It is shown that the elastic modulus for bending mode (i.e. $E_B$) is found as 7~40 GPa and the shear modulus for torsional mode (i.e. $G_T$) is obtained as ~2 GPa, which is comparable to those of Aβ amyloid fibril [43] and also other amyloid fibrils [16] (see Table 1). In addition, the axial elastic modulus (i.e. $Y$) of hIAPP fibril is computed as 12~13 GPa that is comparable to that extracted from MD simulation for Aβ amyloid fibril [12]. It is also shown that the elastic modulus for bending deformation mode, $E_B$, is higher than other elastic moduli for axial stretching and/or torisional deformation modes, i.e. $Y$ and $G_T$. This implies that the dominant deformation for amyloid fibril is the bending deformation rather than axial stretching or torsional deformation modes. This is consistent with our finding that the bending deformation mode corresponds to the lowest frequency normal modes regardless of the length of an amyloid fibril, while other deformation modes such as axial stretching and torsional modes becomes the high-frequency normal modes as the length of a fibril increases (see Fig. 3c). Moreover, it has been remarkably shown that when the length of a fibril reaches ~70 nm, the bending rigidities for two bending deformation modes (i.e. soft mode and stiff mode) becomes identical to each other, which implies that the bending deformation of a fibril obeys the isotropic deformation mode. This may be attributed to the shape of the amyloid fibril (for details, see below). In addition, the bending rigidity of a fibril is strongly dependent on the length of a fibril when its length is less than ~70 nm, whereas the bending rigidity becomes steady-state value when the length reaches ~70 nm. This observed length-dependence mechanical properties have been also found for the case of Aβ amyloid fibril [43], whose bending rigidity becomes steady-state value at the length of 200 nm while the length-dependent bending rigidity is found when the length is less than 200 nm. This suggests that when an aggregated amyloid exhibits the length of ~70 nm, it becomes mechanically stable, that is, it can resist the mechanical bending deformation. In particular, the bending modulus for a hIAPP fibril with its length of 10 nm is 20 GPa, while the bending modulus becomes 40 GPa when the length of a fibril increases to 70 nm. We may conjecture that the critical length scale for the protein aggregation that can toughens the amyloid fibril is ~70 nm. In other words, when protein aggregation leads to the amyloid fibril whose length is ~70 nm, such a fibril becomes mechanically rigid so that it can replace the specific cells such as insulin-secreting β cells at specific sites such as islet of pancreas. We propose that this critical length scale related to the mechanical stability of an amyloid fibril has to be experimentally validated. The length-dependent bending modulus of a fibril has been discussed later (see Section IV. Discussion).

*3. Bending vs. Torsion*

Now, we take into account two deformation modes – bending and torsional deformations – since the torsional rigidity $G_T J$ for torsional deformation mode is quantitatively comparable to the bending rigidity $E_B I$, albeit the torsional mode corresponds to the high-frequency mode. In other words, torsional mode does not significantly contribute to the thermal fluctuation, which implies that the experiments based on the measurement of thermal fluctuation (e.g. Ref. [16]) cannot dictate the mechanical property related to the torsional deformation. In order to quantify the competition between torsional and bending deformations, we introduce the dimensionless quantity, $\eta = E_B I / G_T J$. This dimensionless measure describes the dominant deformation modes between bending and twisting deformations. For instance, a biological fiber such as bacterial flagellar hook and filament exhibits the dimensionless measure $\eta$ such as $\eta < 1$ [57]. This indicates that a bacterial flagellar hook and filament are easily to be bent rather than twisted. However, our finding is very interesting in that, unlike the case of bacterial flagellar hook and filament, the dimensionless measure for an amyloid fibril is $\eta > 1$.

This suggests that amyloid fibril is mechanically resistant to the bending deformation, which implies the mechanical stability of amyloid fibrils with respect to the bending deformation. In particular, as the length of a fibril increases, the dimensionless measure $\eta$ is significantly increased. This implies that the aggregation of β sheet leads to the increase of the mechanical resistance of a fibril to the bending deformation. As shown in Fig. 4, when the length of a fibril reaches ~70 nm, the dimensionless measure $\eta$ approaches ~4, which shows the significant improvement of mechanical resistance to the bending deformation. This value is even much larger than that (e.g. $\eta = 1.3$) for steel solid circular cylinder [58]. This sheds light on the excellent mechanical resistance of an amyloid fibril to the bending deformation that is a significant deformation mode for a biological fiber. This clearly demonstrates how protein aggregation of β sheets gives rise to the mechanical toughening of biological fibril such as amyloid fibril.

*4. Role of Molecular Architecture on the Mechanical Properties*

Because it has recently been found that there are two different molecular architectures for a hIAPP fibril [9, 24], we have studied the role of molecular architecture on the mechanical properties. Even though two molecular architectures, i.e. AHT and PHM, have been experimentally found [9, 24], we consider 4 possible molecular architectures such as PHM, PHT, AHM, and AHT (shown in Fig. 1), respectively, based on a previous study [15]. Fig. 5 depicts the mechanical properties (i.e. elastic bending modulus, axial elastic modulus, and torsional shear modulus) for hIAPP fibrils as a function of their molecular architectures. It is interestingly shown that the elastic bending modulus for hIAPP fibril is in the order of PHT < PHM < AHT < AHM. Specifically, the elastic bending modulus for PHT configuration is ~38 GPa, which is smaller than that (~42 GPa) of AHM configuration at $L = 280$nm. This indicates that β sheets stacked in the antiparallel manner plays a critical role on the mechanical resistance of a fibril to the bending deformation. This is consistent with our finding that the smallest bending rigidity was found for the case of PHT, where four β sheets are stacked in the parallel manner, while the largest bending rigidity was observed for AHM in which four β sheets are stacked in antiparallel manner. The intermediate values for bending rigidity were found for PHM and AHT, where two β sheets are stacked in antiparallel manner but other two β-sheets are stacked in parallel manner. This suggests that the stacking of β sheet (in antiparallel manner) to a fibril is an efficient route to improve the mechanical resistance of a fibril to the bending deformation. This is consistent with a recent finding [18], which reports that β sheet crystal formed by stacking of β strand in antiparallel manner is responsible for its mechanical toughness (stiffness) that determines the excellent mechanical resistance for the spider silk. In order to gain the deep insights into the mechanical properties with respect to the stacking manner, we have also considered the contact order [59] that represents the degree of native contacts formed by hydrogen bonds. It is shown that the β sheets stacked in antiparallel manner (i.e. Fig. 1d) exhibits the contact order of ~0.37, which is larger than that (~0.35) for β sheets stacked in parallel manner (e.g. Fig. 1a). This implies that as the number of hydrogen bond increases (i.e. contact order is increased), it is likely for a fibril to exhibit the higher bending rigidity. In other words, the bending rigidity of a fibril is highly correlated with the degree of native contacts formed by hydrogen bonds. Moreover, it is also found that β sheets stacked in antiparallel manner (e.g. AHM) exhibits the higher torsional rigidity than the β sheets stacked in parallel manner. However, it is provided that the axial stiffness of a fibril can be improved by stacking β sheets in parallel manner rather than antiparallel manner. In other words, when force is applied perpendicular to the plane formed by β sheets, the β-sheet crystal formed by stacking of β sheets in the parallel manner (i.e. parallel strand) is more likely to be mechanically resistant to the force than that formed in the antiparallel manner (i.e. antiparallel strand). This is consistent with a recent finding [60, 61] reporting that parallel strands are mechanically efficient clamp when a force is applied to a single protein domain. This suggests that the mechanical resistance of a fibril is dependent on not only the molecular architecture (i.e. whether it is parallel strand or antiparallel strand) but also the deformation modes (i.e. whether it is bending deformation or axial stretching mode). In case of a fibril, whose important deformation mode is the bending mode, the efficient route to the improvement of mechanical resistance is to stack the β sheets in antiparallel manner. On the other hand, in case of a single protein domain whose significant deformation is the axial stretching mode, the effective way to enhance the mechanical rigidity is to stack the β sheets in parallel manner. This sheds light on how to design the molecular architecture for a biological material that is able to perform the excellent mechanical functions.

*5. Effect of the Twisted Structure*

Our model presumes that the hIAPP fibril exhibits the specific molecular architecture, that is, twisted structure along the fibril axis based on the experimental observation [9]. We conjecture that this twisted structure may be related to the isotropic material properties of an amyloid fibril. As shown in Fig. 3b, when the length of a fibril reaches ~70 nm (corresponding to the helical pitch), the bending rigidity becomes isotropic. In order to investigate the effect of twisted structure on the mechanical properties, we have also considered the molecular structure of a hIAPP fibril such that the structure is assumed to be untwisted. For a fibril whose length is 104 nm, the bending rigidity for a soft axis is $0.7 \times 10^{-26}$ $Nm^2$, whereas the bending rigidity for a stiff axis is $1.1 \times 10^{-26}$ $Nm^2$. Here, two axes (i.e. soft and stiff axes) are shown in Fig. 6. For untwisted structure, even though the bending rigidity for stiff axis is relatively large (i.e. larger than the bending rigidity for twisted AHT configuration), the bending rigidity along the soft axis is smaller than the bending rigidity for twisted AHT configuration. This is consistent with the finite element simulation results on Aβ amyloid fibril [43]. This implies that if an amyloid fibril does not exhibit the twisted molecular structure, the fibril is mechanically weak when the bending deformation occurs along the soft axis. It highlights that the twisted molecular structure is the optimal configuration that leads to homogeneous, isotropic mechanical rigidity for an amyloid fibril. Moreover, the fibril in the untwisted configuration exhibits the torsional shear modulus of 1.87 GPa, which is an order of magnitude lower than that for twisted configuration. This also indicates that the twisted molecular structure is optimal in that it can effectively resist the torsional deformation mode.

## IV. Discussion

In this work, we have studied the mechanical properties of a hIAPP fibril that plays a role in the specific disease expression such as type II diabetes using the coarse-grained normal mode analysis with continuum mechanics theory. It is first shown that the vibrational deformation of a fibril simulated from coarse-grained model can be well described by a continuum mechanics model – Euler-Bernoulli beam theory. It is interestingly found that the bending rigidity of a fibril depends on its length such that the bending rigidity becomes a constant when the length reaches ~70 nm. This length-dependent bending properties cannot be depicted by a conventional Euler-Bernoulli beam theory because Euler-Bernoulli beam theory neglect the shear effects that critically make an impact on the bending rigidity of the beam that possesses low aspect ratio [43].

In order to gain a fundamental insight into the length-dependent bending rigidity, we consider the Timoshenko beam model [62] that accounts for the shear effect on the bending deformation of a beam. This model has been successfully employed for the elucidation of the length-dependent bending behavior of β-sheet crystals [18] and microtubules [51]. When a mechanical force $P$ is applied perpendicular to the longitudinal direction of a fibril, the total bending deflection of a fibril consists of the deflections due to bending deformation and shear deformation, respectively.

$$\delta = \delta_B + \delta_S = \frac{PL^3}{aE_B^0 I} + \frac{cPL}{bG_S A} \quad (3)$$

where $\delta$ is the total bending deflection, $\delta_B$ is the deflection due to bending deformation, $\delta_S$ is the deflection due to shear deformation, $E_B^0$ is the length-independent bending elastic modulus,

$G_S$ is the intrinsic shear modulus, $I$, $A$, and $L$ represent the cross-sectional moment of inertia, the cross-sectional area, and the length, respectively, of an amyloid fibril, $a$ and $b$ are constants dependent on the boundary condition, and $c$ is a shear coefficient that depends on the cross-sectional shape, e.g. $c = 1.5$ for rectangular cross-sectional shape. Because we have a relation between bending deflection and effective bending rigidity, i.e. $\delta = PL^3/aE_BI$, the effective elastic bending modulus, $E_B$, can be represented in the form

$$E_B = E_B^0 \left(1 + \frac{a}{b}\frac{cE_B^0 I}{G_S A L^2}\right)^{-1} \qquad (4)$$

Eq. (4) clearly demonstrates the dependence of the bending elastic modulus on the length of a fibril. The Timoshenko beam theory depicted in Eq. (4) is well fitted to the relationship between the bending rigidity and the length of a fibril (Fig. 3b), which is computationally obtained from our simulation based on coarse-grained normal mode analysis. In particular, it is found that the hIAPP fibril exhibits the intrinsic properties such as $E_B^0 I = 0.8 \times 10^{-26}$ Nm$^2$ and $G_S = 1.1$ GPa. In order to quantify the shear effect on the bending deformation of an amyloid fibril, we introduce the dimensionless measure $\psi$ defined as $\psi = (a/b)(cE_B^0 I/G_S A L^2)$, which represents how the bending deformation is attributed to the shear effect. Based on this dimensionless measure with intrinsic properties, it is interestingly shown that when the length of a fibril becomes ~70 nm, the dimensionless quantity $\psi$ becomes less than 1%, which indicates that shear effect does not play any role in the bending deformation of a fibril. As described earlier, the length scale at which shear effect plays an insignificant role is identical to the critical length scale at which the bending rigidity of a fibril becomes a constant regardless of the length. This highlights that the length-dependent mechanical stability of an amyloid fibril may be highly correlated with the shear effects on the bending deformation. For instance, when the length of a fibril is ~10 nm, the dimensionless quantity $\psi$ becomes ~60 %, which significantly reduces the bending rigidity of an amyloid fibril. It is implied that the critical length scale, at which the amyloid fibril becomes mechanically stable, is determined from the intrinsic properties of a fibril such as its length-independent bending elastic modulus $E_B^0$ and intrinsic shear modulus $G_S$.

Moreover, the dependence of the bending properties on the length has been observed regardless of types of molecular architecture – i.e. whether the fibril is formed based on stacking in parallel or antiparallel manner, and whether the fibril exhibits twisted or untwisted structure. In particular, for all 4 possible configurations (i.e. PHM, PHT, AHM, and AHT), the critical length scale at which the bending rigidity becomes a constant is ~70 nm, which corresponds to the length scale at which the shear effect does not play any role in the bending deformation. Moreover, for untwisted molecular structure, the anisotropic bending rigidity is also dependent on the length of a fibril. In particular, when the length of a fibril is in the range of 50 nm to ~70 nm, both stiff and soft bending rigidities becomes steady state. This length scale corresponds to the case in which the shear effect becomes unimportant in the bending deformation. This clearly elucidates how the mechanical properties, particularly bending properties, of an amyloid fibril are determined based on the shear effect.

Our simulation results have the implications for the future applications such as molecular therapeutics. Specifically, as suggested in Ref. [9, 24], the different physiological conditions lead to hIAPP fibril's different molecular architectures such as AHT and PHM configurations. As shown in our simulation results, the mechanical properties (e.g. bending modulus) of these two configurations are quite different from each other. As the consequence, if the relationship between the physiological condition and the fibril's molecular architecture that encodes the material properties is unveiled, then it is straightforward to induce the changes in the material properties of fibrils through the changes in physiological conditions, which gives rise to the degradation of the fibril. This implies that our simulation may open a new, efficient route to the future therapeutics such that our simulations provides the relationship between the fibril's material properties and the fibril's molecular architecture that can be affected by the change in the physiological conditions, as long the correlation between the physiological condition and the fibril's molecular architecture is available. Moreover, as reported in recent literatures [11, 63, 64], the mutation of a couple of single amino acids in the fibril makes a critical effect on both the molecular architecture and mechanical properties of protein filaments such as collagen fiber [64] and Aβ amyloid fibril [11]. This indicates that the simulation can provide the fundamental insights into how to degrade the mechanically stable amyloid fibril through the introduction of mutations. It has to be noted that our model in its current form lacks the atomistic details so that our current model is unable to capture the mutation-induced changes in the mechanical properties of protein materials such as amyloids. For the improvement of our model in order to capture the mutation-induced degradation of fibrils, our model has to be appropriately modified by consideration of all atoms (or more refined coarse-grained beads) with chemical bond-dependent force constants (e.g. atomic elastic network [65]), which may enable the predictions on the amino acid sequence-dependent mechanical properties, and consequently the mutation-driven degradation of fibrils.

In conclusion, we have shown how the mechanical properties (particularly, bending properties) of amyloid fibrils can be optimized. Specifically, as shown in our work, amyloid fibril can exhibit the excellent bending rigidity through the design of the configuration of hydrogen bonds between β sheets. The excellent bending resistance for a fibril is attributed to the antiparallel strands for which the number of hydrogen bonds (i.e. contact order) is optimized. Moreover, we have also found that the twisted molecular structure is responsible for not only the isotropic bending properties of a fibril but also its mechanical stability. In particular, if a fibril exhibits the untwisted molecular structure, the bending rigidity for soft axis is reduced in comparison with that for twisted structure of a fibril. More remarkably, it is shown that the length-dependent bending properties of a fibril can be depicted by a continuum mechanics theory such as Timoshenko beam theory. This suggests that the critical length scale at which the bending rigidity becomes a steady state is determined in such a way that the shear effect comes to play an unimportant role in the bending deformation.

**Acknowledgement**

K.E. appreciates the financial support from National Research Foundation of Korea (NRF) under Grant No. NRF-2010-0026223. S.N. gratefully acknowledges the financial support from NRF under Grant No. NRF-2010-0001642 and NRF-2008-314-D00012

# Figure Captions

**Fig. 1.** Four possible configurations for cross-β that composes the hIAPP fibril: **a.** Parallel Homo (PHM), **b.** Parallel Hetero (PHT), **c.** Antiparallel Homo (AHM), and **d.** Antiparallel hetero (AHT). Here, the interspacing distance between $\beta_1$ and $\beta_2$ is given as 10.5 Å, and the distance between $\beta_1$ and $\beta_3$ is given by 4.87 Å.

**Fig. 2. a.** Four fundamental deformation modes (i.e. bending, torsional, and axial stretching modes) of a hIAPP fibril, whose length is equal to the one helical pitch, and the structure for which is AHT. Soft bending, stiff bending, torsional, and axial stretching modes are presented from the left panel to the right panel. The solid black lines for bending modes represent the bending deformation modes obtained from a continuum mechanics theory (i.e. Euler-Bernoulli beam model). **b.** The natural frequencies for each deformation mode are presented. The filled symbols show our simulation results on hIAPP fibril, while the empty symbols indicates the simulation results on the Aβ amyloid fibril whose length is 19.28 nm reported by Xu *et al*. [43].

**Fig. 3. a.** Elastic moduli for four deformation modes (i.e. bending, torsional, and axial stretching modes) are shown as a function of the length of a fibril. **b.** Bending rigidities (for both soft and stiff bending modes) are presented with respect to the length of a fibril. **c.** Mode indices for each deformation mechanism are shown as a function of the length of a fibril.

**Fig. 4.** Dimensionless measure, $\eta = E_B I / G_T J$, (which represents the dominant deformation modes between bending and torsional modes) as a function of the length of a fibril.

**Fig. 5.** Mechanical properties of the hIAPP fibril are presented as a function of not only the molecular architectures but also the length of a fibril: **a.** The bending elastic modulus (for soft mode), **b.** the torsional shear modulus, and **c.** the axial elastic modulus are computed for the hIAPP fibril whose conformations are PHM, PHT, AHT, and AHM, respectively.

**Fig. 6.** Bending rigidities (for both soft and stiff modes) for hIAPP fibril, which exhibits the untwisted molecular structure, are presented.

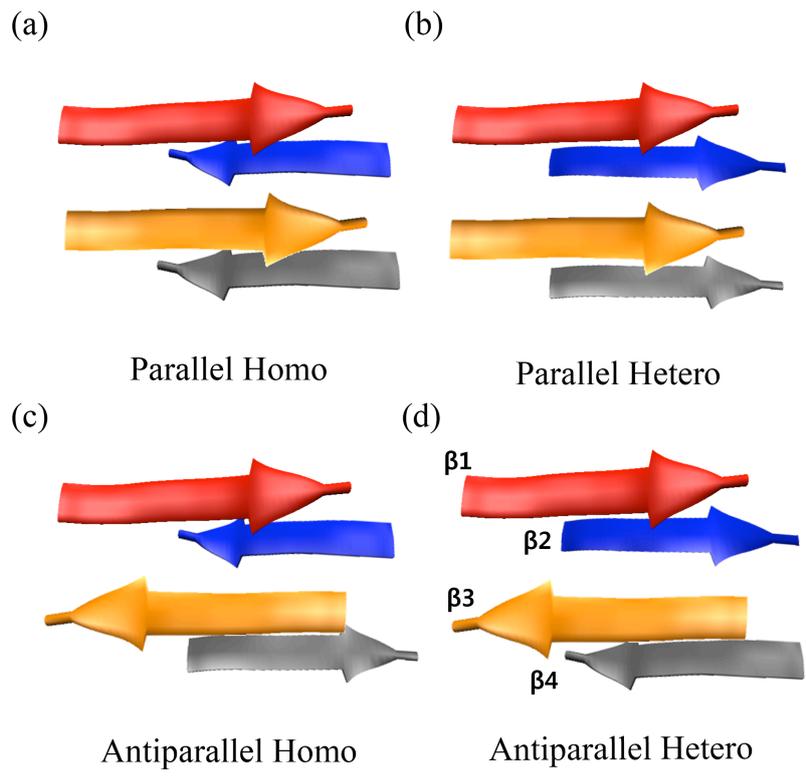

Figure 1

(a)

L=17.045nm

Soft bending 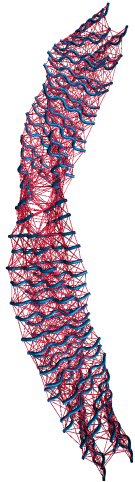  Stiff bending 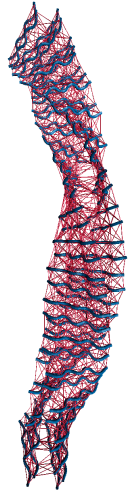  Torsion 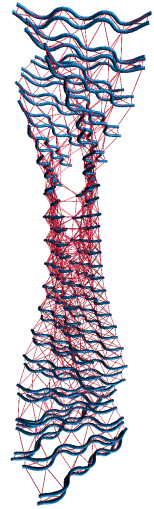  Axial 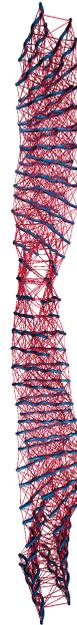

(b)

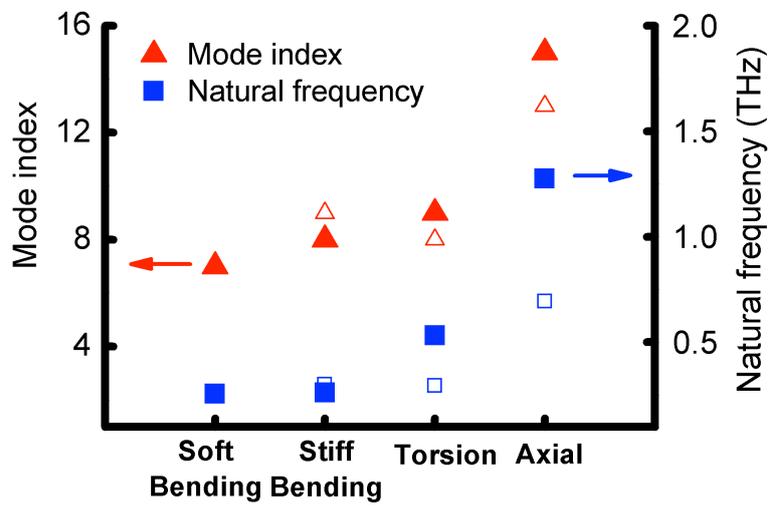

Figure 2

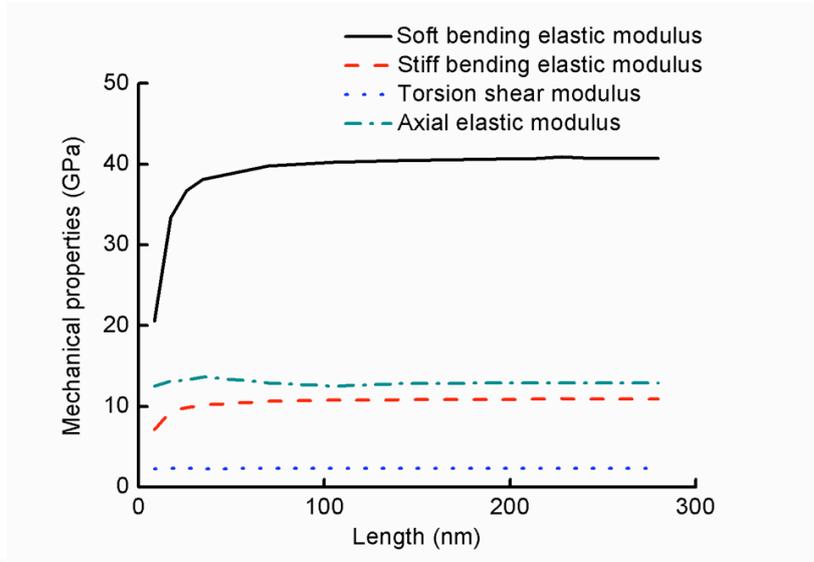

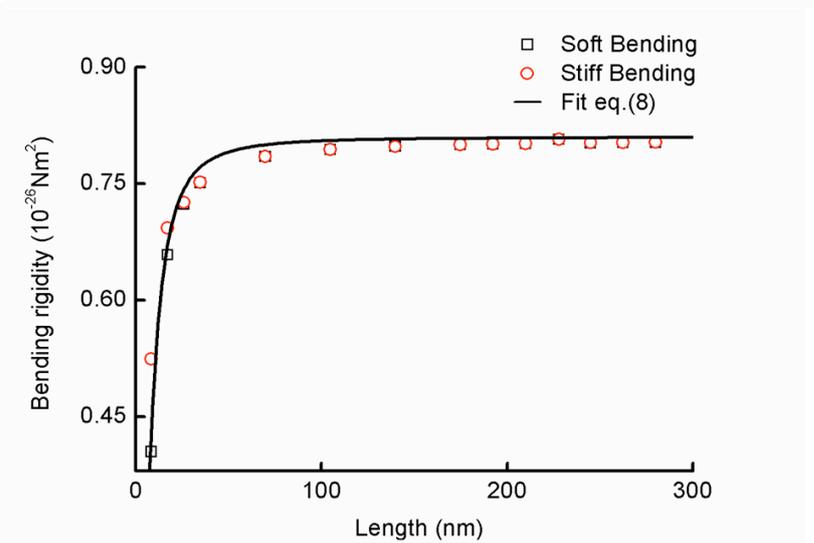

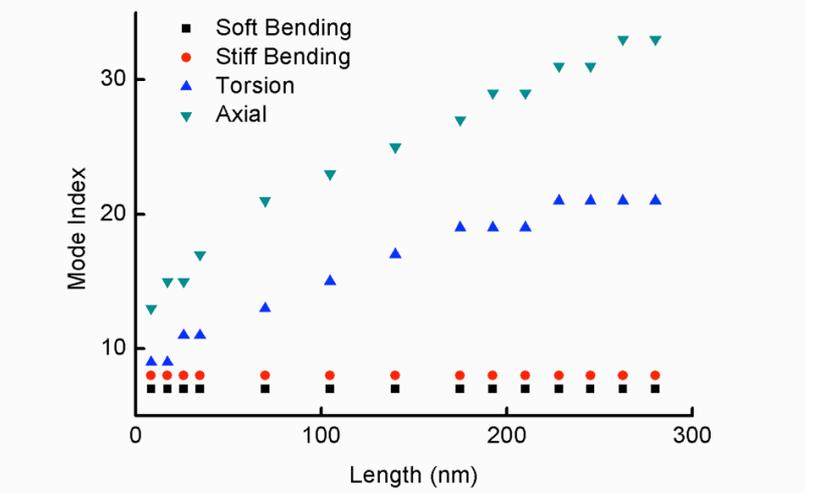

Figure 3

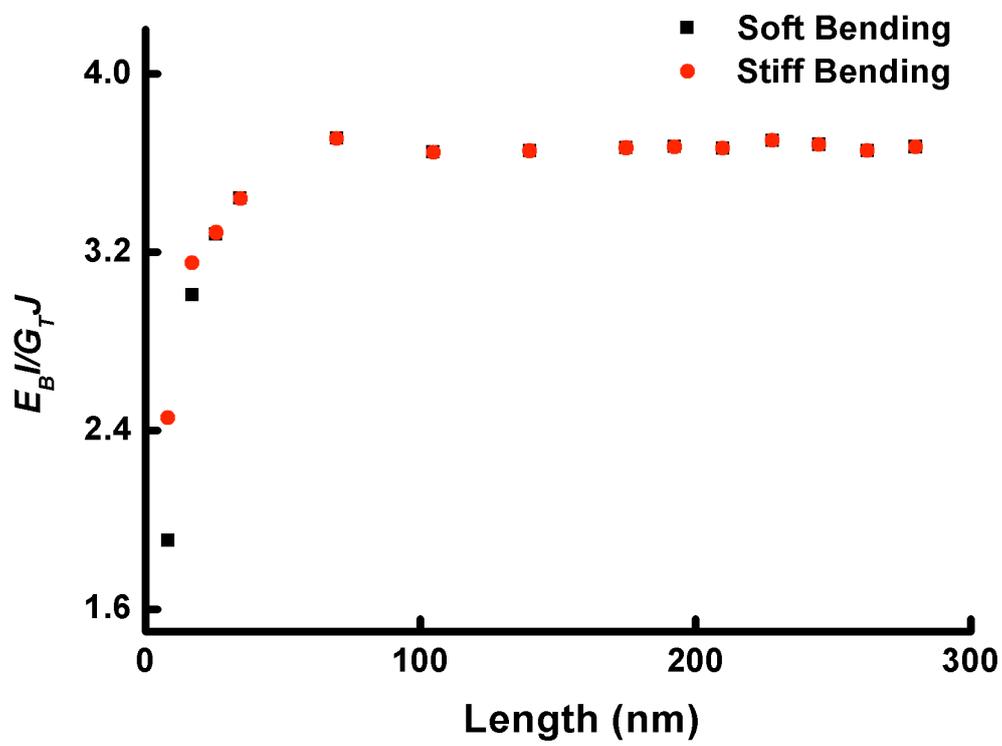

Figure 4

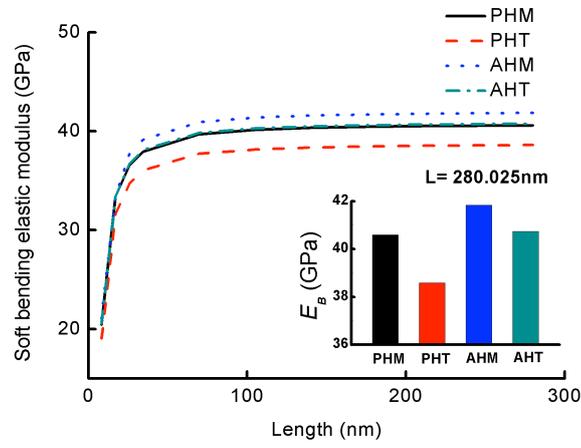

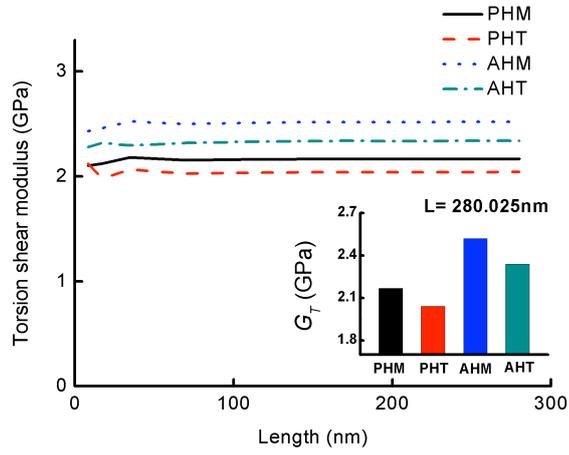

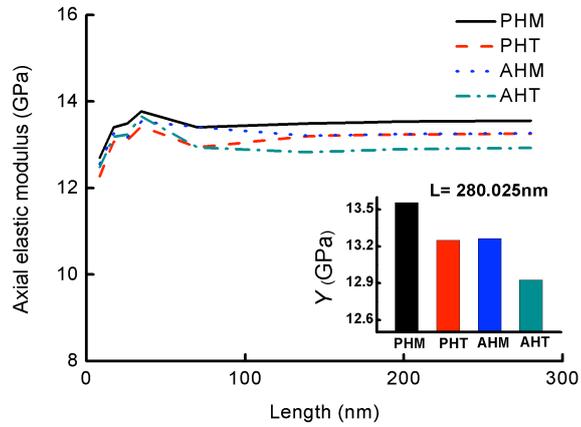

Figure 5

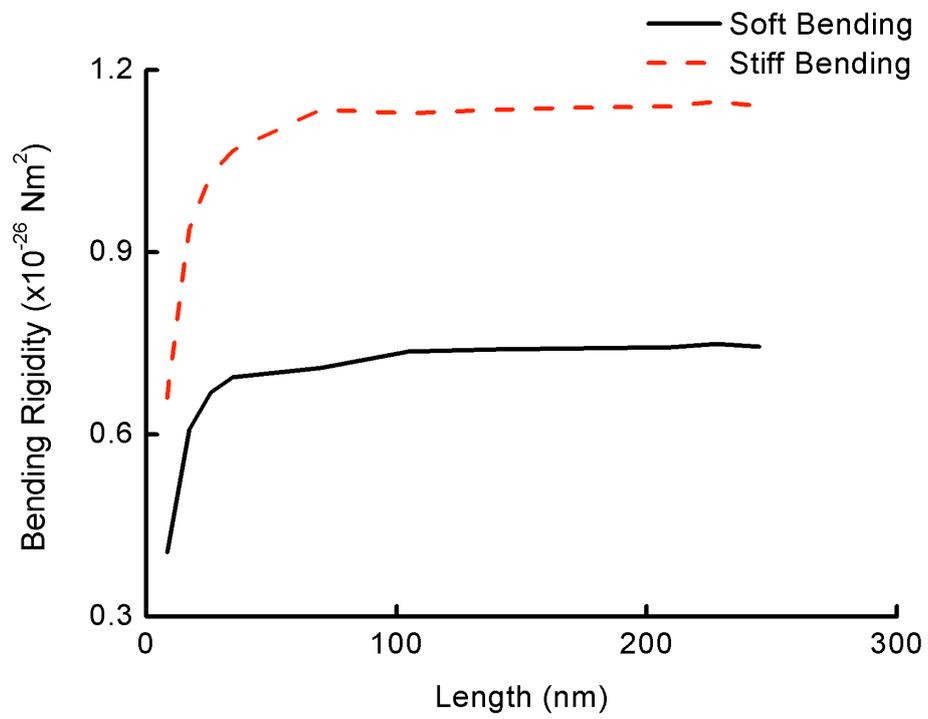

Figure 6